\documentclass[doublecol]{epl2}

\usepackage{setspace}

\usepackage{graphicx}
\usepackage{dcolumn}% Align table columns on decimal point
\usepackage{bm}% bold math
\usepackage{epsfig,amsmath, amsfonts, amssymb, xcolor,hyperref}
\usepackage{ stmaryrd}
\usepackage{soul}
\usepackage{textcomp}

\begin{document}

\title{Surface states and the thermal Casimir interaction }
\date{\today}
\author{Douglas B. Abraham\inst{1,2,3} \and
  Anna Macio\l ek\inst{3,4}}
\shortauthor{A. Macio\l ek \etal}
\institute{
  \inst{1}  Theoretical Physics, Department of Physics, University of Oxford, 1 Keble Road, Oxford OX1 3NP, United Kingdom \\
  \inst{2} Centre for Non-linear Studies, Los Alamos National Laboratory, Los Alamos, NM 87545, USA \\
  \inst{3} Max-Planck-Institut f{\"u}r Intelligente Systeme,
  Heisenbergstra\ss e~3, D-70569 Stuttgart, Germany\\
  \inst{4} Institute of Physical Chemistry, Polish Academy of Sciences,
  Department III, Kasprzaka 44/52, PL-01-224 Warsaw, Poland 
}
\date{today}

\abstract{
 Using exact calculations, we elucidate the significance of the surface (bound) states for the thermal Casimir interactions for
an Ising strip with a finite width.
The surface state arises  whenever an imaginary wavenumber mode appears in the spectrum of the transfer matrix,
taken in the direction  parallel to the edges of the strip. 
Depending on the boundary conditions, the imaginary modes emerge  below or above the bulk critical temperature, or below 
the wetting temperature  of a single surface with surface magnetic field. The bound states  are responsible for the
  strong asymmetry of the Casimir forces between the super- and sub-critical regimes and for  their sign. 
Our analysis uses the fact that  the Casimir forces have  two  mathematical forms. We show that these very different
representation are the same and in the process find the origin of the asymmetry.
  }


\maketitle

It has been known for about half a century that the existence of quantum fluctuations in the electromagnetic vacuum is revealed 
by the existence of long-ranged, largely attractive, forces between objects  \cite{Casimir}. For instance, in the geometry of parallel plates at a distance $L$
apart, the wavenumbers in the direction  perpendicular to the plates become discrete and this produces
 a change in the energy, which (per unit area) behaves as
$ 
\delta E(L)\propto \hbar c/L^3
$
for sufficiently large distances, $c$ is the speed of the light and $\hbar$ is a Planck's constant.
Fisher and de Gennes \cite{FdG} made the crucial observation that in condensed matter systems, say uniaxial classical ferromagnets
 and their analogues, the confining surfaces restrict  order parameter fluctuations leading to to "entropic"
forces between the surfaces. Scaling theory applied to these fluctuations predicts the existence of forces per unit area 
in the critical region (at zero bulk field) of the form  (in the following
all free energies and forces are expressed in units of $k_BT$)
\begin{equation}
\label{eq:1}
 {\cal F}_{Cas} =  L^{-d}\Theta(sgn(T-T_c)L/\xi)
\end{equation}
where $T_c$ is the bulk critical temperature,  $\xi$ is the bulk correlation length and the distance between the confining surfaces is
 $L$ \cite{FdG}.
These forces are of significance  on the micro- and nanoscale as shown in experiments on wetting films
and colloidal systems \cite{pershan,beysens,nature,bonn}. 

The finite-size scaling function $\Theta$ is  expected to 
depend on the geometry and  on the boundary conditions (BCs)  imposed by the confining surfaces.
Mathematical consistency may allow small and large argument behavior in (\ref{eq:1}) to be inferred, but to go further, detailed calculations of $\Theta$ 
are needed.
The main generally applicable techniques are Monte-Carlo simulations \cite{vas-08} and the de Gennes-Fisher-Upton local functional method \cite{upton}. The success of the first depends crucially on a thorough
understanding of finite-size effects; the second is phenomenological in character, but with the benefit of
containing no adjustable parameters. In view of these
matters, exactly solvable models have a special relevance, since they afford insights not otherwise available.

In this Letter, we shall consider Casimir forces in the planar Ising model with a strip geometry.
This encompasses binary mixtures and lattice gases, with the surface magnetic fields playing the role  of surface (differential) 
fugacities.  Depending on the method of calculation used
to obtain them, the results of the Casimir force 
come in two very different {\it mathematical} forms. One of them, a {\it mode sum}, has received little analytical attention so far.
 It has two advantages:
firstly, the finite geometry imposes the discretisation condition in this mode sum, as in the quantum Casimir case. 
The second advantage, and this is one of the results of this Letter, is that the strong asymmetry  between the super-
and sub-critical regimes receives for the first time a {\it microscopic} interpretation
 through the contribution of bound states, or surface modes.
Their  existence in this context has been known 
for quite some time \cite{abraham}, but not the interpretation, which we provide  in this Letter   in terms of 
the Fisher-Privman finite-size scaling theory \cite{privmann_fisher}.
The other representation of the Casimir force for strips, this time   as an {\it integral},
 comes either from Au Yang and Fisher \cite{AuYang_Fisher,evans_stecki}
or from  extension of the  Schultz,  Mattis  and Lieb (SML)  method \cite{sml,Roya}, with  the following advantages: 
it is ideal for taking the scaling limit, and also for investigating real analyticity in $x$ of $\Theta(x)$  from (\ref{eq:2}). 
Further, although it is the obvious choice for numerical work, it gives no insight into the striking asymmetry mentioned above 
\cite{dbaam}.
The most recent contribution along this line explains how the boundary fields can be adjusted to produce both attractive and repulsive critical
Casimir interactions \cite{dbaam}.

\begin{figure}
\onefigure[width=0.45\textwidth]{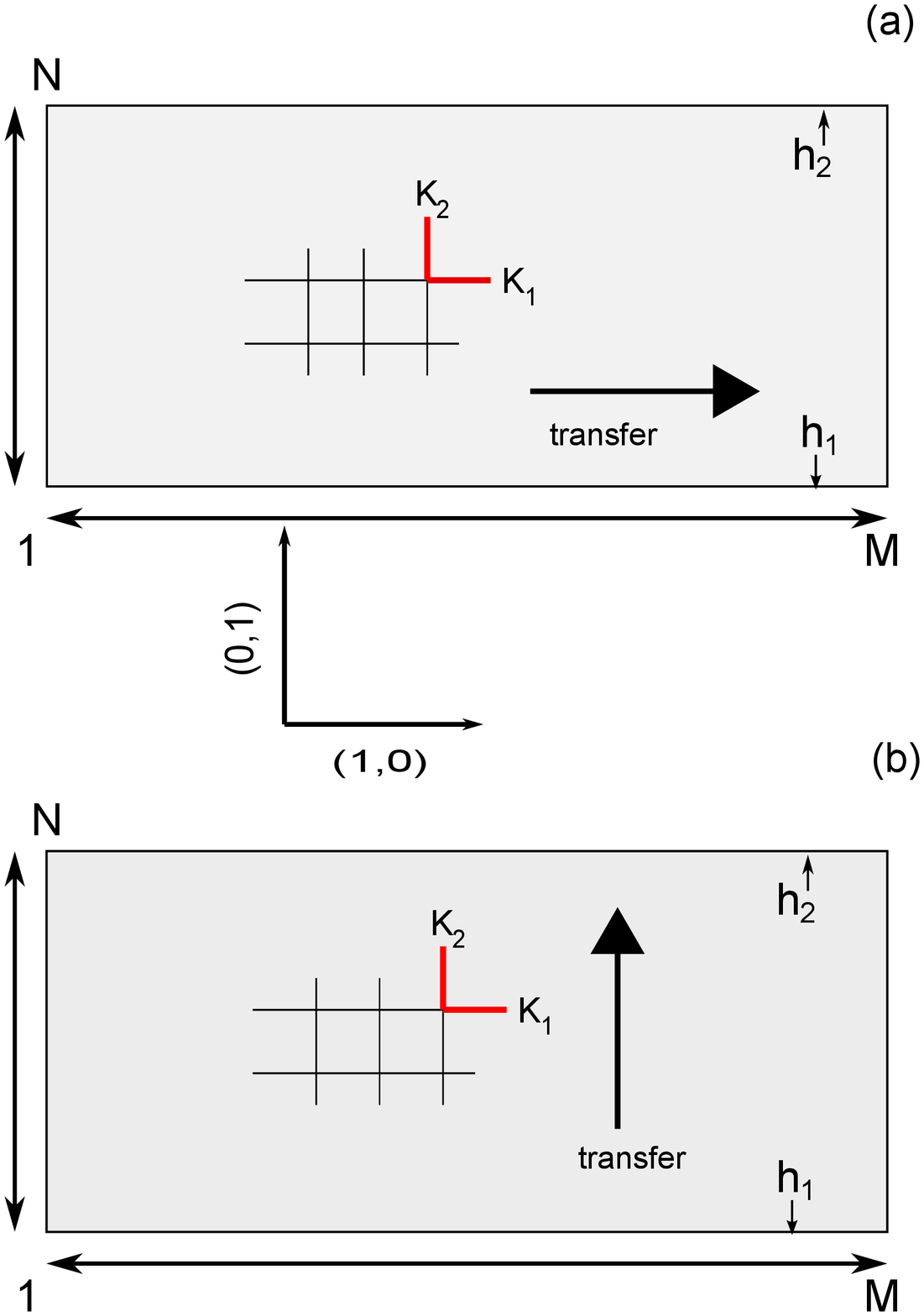}
  \caption{(color online)  An Ising strip with  $M$ columns and $N$ rows. We are interested in the limit $M\to \infty$ with $N<\infty$.
Surface fields $h_1$ and $h_2$ are applied at the  bottom and top edges; in the present paper we consider free boundaries 
$h_1=h_2=0$ and the $(+,+)$ boundary conditions $h_1=h_2=\infty$. 
In the model (a) the transfer matrix  is along  the $(1,0)$ axis whereas in the model (b) 
along the $(0,1)$ direction.}
  \label{fig:0}
\end{figure}

We now specify the model more precisely. As shown in Fig.~\ref{fig:0}, we have an Ising strip with cyclic BCs in the $(1,0)$ direction.
The energy for a configuration $\{\sigma\}$ of spins is 
\begin{eqnarray}
\label{eq:2}
 E_{N,M}(\{\sigma\})=-\sum_{m=1}^M\sum_{n=1}^{N}K_1\sigma_{m,n}\sigma_{m+1,n} \\ \nonumber 
-\sum_{m=1}^M\sum_{n=1}^{N-1}K_2\sigma_{m,n}\sigma_{m,n+1}+E_{boundary}.
\end{eqnarray}
The term $E_{boundary}$ just concerns the rows $n=1$ and $n=N$. This situation can be handled by a transfer matrix (TM) 
building the lattice  in the $(0,1)$ 
direction, where the matrix has cyclic symmetry in the $(1,0)$ direction;  the SML technique \cite{sml} may be applied. 
The alternative is to transfer in the $(1,0)$ direction; cyclic symmetry no longer 
obtains, but the TM spectrum can still be found \cite{abraham,abraham_martin,maciolek,maciolek_stecki} for the BCs which we consider.

\vspace*{0.5cm}

The free energy per unit length in the $(1,0)$ direction, as derived from the transfer in the $(0,1)$ direction  is  \cite{dbaam}
\begin{eqnarray}
 \label{eq:3}
f(N)=-\frac{1}{4\pi}\int_{-\pi}^{\pi}d\omega \ln\Bigg[ A_1A_2e^{N\hat\gamma(\omega)} 
% \nonumber \\
+ B_1B_2e^{-N\hat\gamma(\omega)}\Bigg]
\end{eqnarray}
up to an additional term $(N/2)\ln(\sinh 2K_2)$. Here, $\hat\gamma(\omega)$ 
is the Onsager function   given  by \cite{O44} 
\begin{equation}
\label{eq:4}
\cosh \hat\gamma=\cosh 2K_1\cosh 2K^*_2 - \sinh 2K_1\sinh 2K^*_2\cos \omega, 
\end{equation}
where the non-negative real branch is taken for real $\omega$,  $K_j^*$ is the dual coupling given
 by the involution $\sinh 2K_j\sinh 2K_j^*=1, j=1,2$;
$A_j(\omega)$ and $B_j(\omega)$  depend on the BCs. 
For free boundaries, we have $A_1=A_2=e^{\hat\gamma}\cos(\hat\delta'/2)$ and
$B_1=B_2=e^{-\hat\gamma}\sin(\hat\delta'/2)$, for $(+,+)$ BCs with all spins on the bottom and top boundary fixed at the value $+1$,
 we have
$A_1=A_2=e^{\hat\gamma}\cos(\hat\delta^*/2)$ and $B_1=B_2=e^{-\hat\gamma}\sin(\hat\delta^*/2)$. The angles $\hat\delta^*$ 
 and $\hat\delta'$  are elements of the Onsager hyperbolic triangle \cite{O44}, given by the formula
$\sin\hat\delta'(\omega)=\sinh 2K_1\sin(\omega)/\sinh\hat\gamma(\omega)$ and 
$\cos\hat\delta'=(\cosh2K^*_2\cosh\hat\gamma-\cosh2K_1)/(\sinh 2K^*_2\sinh\hat\gamma)$.
The angle $\hat\delta^*(\omega)$ has $K_1^*$ and $K_2$ interchanged.
  Extracting  from (\ref{eq:3})  the bulk contribution
  $Nf_{\infty}=-N(1/4\pi)\int_{0}^{2\pi}\hat\gamma(\omega)d\omega$, which  is easy,
and an additional,  $N$ {\it independent, surface} contribution 
$f_s=-(1/4\pi)\int_{0}^{2\pi}\log\left[A_1A_2\right]d\omega$ \\
gives the Casimir free energy:
\begin{equation}
\label{eq:5}
{\cal F}_{Cas}=-\frac{1}{4\pi}\int_{-\pi}^{\pi}d\omega\ln\left(1+e^{-2N{\hat\gamma}(\omega)}\tan^2\left(\hat\delta(\omega)/2\right)\right),
\end{equation}
where  $\hat\delta(\omega)=\hat\delta_1(\omega)+\hat\delta_2(\omega)$, and 
$\hat\delta_i(\omega), i=1,2$ depend on the BCs. 
For free boundaries $\hat\delta(\omega)=\hat\delta^*(\omega)$, whereas for $(+,+)$ BCs 
$\hat\delta(\omega)=\hat\delta'(\omega)$.

On the other hand, the transfer matrix  $V$ for the $(1,0)$ direction has the diagonal form \cite{abraham,abraham_martin,maciolek,maciolek_stecki}
$V=\exp\left[-\frac{1}{2}\sum_{j=1}^N\gamma(\omega_j(N))(2X_j^{\dagger}X_j-1)\right]$. 
The $X_j$ are Fermion operators with anticommutation relations:
 $\left[X_j,X_l\right]_+=0, \left[X_j,X_l^{\dagger}\right]_+=\delta_{jl}$. Thus 
$2X_j^{\dagger}X_j-1$ has  eigenvalues $\pm 1$.
They are a linear combinations of the spinor operators $\Gamma_n$ \cite{sml}:
\begin{equation}
\label{eq:6}
X^{\dagger}(\omega_j)=\sum_{n=1}^{2N}y_n(\omega_j)\Gamma_n,
\end{equation}
where functions $y_n(\omega)$ are known, e.g. for free boundaries see Ref.~\cite{abraham}.
If $x$ is the ``ordering'' direction in a sense that the Pauli matrix $\sigma^x_n$ \cite{sml} measures the spin at site $n$, then 
\begin{equation}
\label{eq:6a}
\Gamma_{2n-1}=P_{n-1}\sigma^x_n \quad \Gamma_{2n}=P_{n-1}\sigma^y_n
\end{equation}
where $P_0=1$  and $P_n=\prod_{j=1}^{n}(-\sigma^z_j), j=1,\ldots, N$ and $\sigma^i_n$ are Pauli operators \cite{sml}.
In the customery language, $P_n$ is a string of disordering operators because
 $(-\sigma_j^z)$ reverses the spin at site  $j$.
We emphasise that this method does not give the $X(\omega_j)$ vaccum $\mid\Phi\rangle$ 
(the maximal eigenvector, since $\gamma(\omega)\ge 0$)
{\it constructively}, unlike the SML case \cite{sml}.
The  maximum   eigenvalue of $V$ is $\Lambda_{max}(N)=\exp (1/2)\sum_{j=1}^N\gamma(\omega_j(N))$
and 
the free energy  $f(N)$ per unit length in the $(1,0)$  direction 
is now
\begin{equation}
\label{eq:7}
f(N)=-\frac{1}{2}\sum_{j=1}^N \gamma(\omega_j(N)),
\end{equation}
valid for $h_1h_2\ge 0$. This is evidently quite unlike (\ref{eq:3}) in form.
The function $\gamma(\omega)$ is as $\hat\gamma(\omega)$ in (\ref{eq:4}), but with $K_1$ and $K_2$ interchanged. 
The $\omega_j(N), j=1,\ldots, N$ are the solutions of the 
discretisation equation imposed by the strip geometry
\begin{equation}
\label{eq:8}
\exp 2iN\omega=\exp i\left(\delta_1(\omega)+\delta_2(\omega)\right).
\end{equation}
In this approach, the interpretation of the angles   $\delta_1$ and $\delta_2$ is clear; they are phase shifts
for reflection of lattice fermions from the surfaces and (\ref{eq:7}) is a consistency condition requiered by the presence 
of two such bounding surfaces at a {\it finite} separation $N$ \cite{abraham,abraham_martin}.
The functions $\delta(\omega), \delta_j(\omega), j=1,2$ are given by the same expressions as $\hat\delta(\omega),\hat\delta_j(\omega), j=1,2$,
but with $K_1$ and $K_2$ interchanged. 

With free boundary, a  study  of the solutions of (\ref{eq:7}) and of the 
non-triviality of the eigenvectors 
which relate to them mandates that we take the solutions in $(0,\pi)$.
In fact, there is one solution in each interval $[\pi(j-1)/N,\pi j/N], j=1,\ldots, N$ provided $K_1^*>K_2, (T>T_c)$. But for 
$K_1^*<K_2, (T<T_c)$, there is a solution as above with $j=2, \ldots, N$. If $j=1$, there is such a solution but only if $N<(\delta^*)'(0)$.
If we take the pair $-\omega_1(N), \omega_1(N)$,  then as we pass
from $N<(\delta^*)'(0)$ to $N>(\delta^*)'(0)$, there is a {\it bifurcation} to a pair of imaginary 
solutions $\omega=\pm iv(N)$, 
one of which is to be taken in the solution set. For $v(N)>0$, up to order $\exp(-4N\gamma(0))$ we have
\begin{equation}
\label{eq:9}
v(N)\simeq \hat\gamma(0)- 2\left(\sinh 2K_2\sinh 2K_1\right)^{-1}\sinh\hat\gamma(0)e^{-2N\hat\gamma(0)},
\end{equation}
with $\hat\gamma(0)=2(K_1-K_2^*)$ (see Eq.~(\ref{eq:4})). The corresponding  $\gamma(iv(N))$ is
given up to order $\exp(-2N\gamma(0))$    
\begin{equation}
\label{eq:10}
\gamma(iv(N))\simeq 2\sinh 2K_1^*\sinh\hat\gamma(0)e^{-N\hat\gamma(0)},
\end{equation}
which clearly demonstrates asymptotic degeneracy as $N\to\infty$ \cite{abraham,kac,lassetre}.
The  bifurcation of a solution corresponding to $j=1$ 
  is a mathematical manifestation of the shift of a pseudocritical temperature in a film
with respect to the bulk critical temperature $T_c$. Recall that in a confined two-dimensional Ising system there is no true ordering phase transition; it becomes exponentially rounded
and  shifted towards lower temperatures.
 Such a shift, which is  $\sim (1/N)$, 
 has been predicted from a  scaling
analysis by Nakanishi and Fisher \cite{nakanishi_fisher}.

In order to  extract the Casimir free energy from (\ref{eq:6}),
we use a contour integral method with a suitable summation kernel.
The first use of a summation kernel in problems of this type, somewhat tangentially,
 is in Onsager's calculation of surface tension \cite{O44}. There, the wavenumbers are equally spaced. 
Further, there are applications in stochastic processes \cite{NW}, and, moreover, the
 general problem was considered in \cite{abraham}, but in a different context.
The contour integral is 
\begin{equation}
\label{eq:11}
\frac{1}{2}\sum_{j=1}^N \gamma(\omega_j(N))= \frac{1}{8\pi i}\oint_Cd\omega\gamma(\omega)\frac{1}{F_N(\omega)}\frac{dF_N}{d\omega}+R
\end{equation}
where $R=-(1/2)(\gamma(0)+\gamma(\pi))$, with
summation kernel
$F_N(\omega)=e^{2iN\omega}-e^{i\left(\delta(\omega)\right)}$; the integration
 contour $C$ surrounds every   zero of $F_N(\omega)$   in the strip  
$-\pi< Re \omega <\pi$, with special attention to the behavior 
at $\omega=\pm\pi$ and $\omega=0$. 
None of these is  an allowed value, since the associated eigenfunction is trivial.
The terms $\gamma(0)$ and $\gamma(\pi)$ do not report in the thermal Casimir force
as they are $N$- independent. The development of the integral in (\ref{eq:11}) 
will be given elswhere. After extracting the bulk term, we find that the rather
different representation are indeed the same. Since (\ref{eq:5}) has a scaling 
limit (by direct construction), so does the Casimir free energy in the equivalent form
\begin{equation}
\label{eq:12}
f_{Cas}(N)=
-\frac{1}{2}\sum_{j=1}^N \gamma(\omega_j(N))+\frac{N}{4\pi}\int_{-\pi}^{\pi}\gamma(\omega)d\omega
\end{equation}
a result which would be difficult to extract from a direct assault on (\ref{eq:5}). We wish to assess the contribution 
of a single term in the sum to (\ref{eq:12}). Provided this has a scaling limit, then it is meaningful to compare
it with (\ref{eq:5}).

\begin{figure}
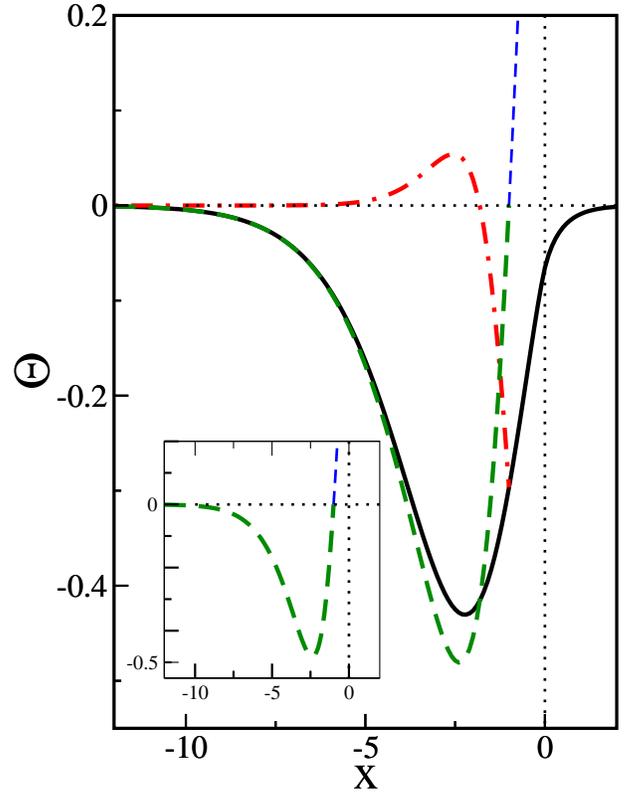

\onefigure[width=0.45\textwidth]{cs_substr_free.eps}
  \caption{(color online) The  surface mode contribution to the scaling function of the critical Casimir force  
$\Theta^{c}=N^2{\cal F}_{Cas}^{(c)}$
for the isotropic ($K_1=K_2$) Ising strip with free boundary conditions (dashed green line) shown together with 
the  total  scaling function $\Theta(x)$ (solid black line). The red  dot-dash line is the result of the substraction of 
 the part coming from the imaginary wave number
$k=iv(N)$ from the $\Theta(x)$. The surface mode contribution (the dashed green line) is equal to 0  at $x=-1$. The dashed blue line shows the contribution to the scaling function of the critical Casimir force coming from the real wave number $\omega_1$ for $-1<x<0$.  }
  \label{fig:1}
\end{figure}

 In particular, it is pertinent to compare the contribution of $(1/2)\gamma(iv(N))$  coming
from a {\it c}omplex mode to the Casimir force
in the scaling limit as calculated numerically from  (\ref{eq:5})  (see Eq.~(3) in Ref.~\cite{dbaam}). 
This is shown in Fig.~\ref{fig:1} for $K_1=K_2$ and free boundaries. The Casimir force is the negative of derivative of the Casimir free energy  
with respect to $N$.
Taking the scaling limit as $N\to \infty$, $\hat\gamma(0)\to 0$, such that $x=N\hat\gamma(0)sgn(T-T_c)$ is fixed 
of the contribution to the Casimir force from the 
imaginary mode, one obtains 
\begin{equation}
 \label{eq:13}
{\cal F}_{Cas}^{c}= -\frac{\sinh2K_1^*(T_c)}{2}\frac{1}{N^2}\frac{z^2\sqrt{x^2-z^2}}{\left(|x|+x^2-z^2\right)},
\end{equation}
for $x<0$.
The function  $z(x)$ is calculated from
\begin{equation}
 \label{eq:14}
\exp(-2z)=(x+z)/(x-z).
\end{equation}
For $x<-1$ there is one solution, for which
the denominator of (\ref{eq:13}) is strictly positive. For $-1<x<0$ ( and incidently for $x>0$ )
(\ref{eq:14}) has no real solution.
As one can see from  Fig.~\ref{fig:1} the imaginary wave number yields  the dominant
contribution to the scaling function $\Theta(x)$ of the critical Casimir force
 which is negative.  Moreover, its magnitude is larger than that of 
$\Theta$.
From  Fig.~\ref{fig:1}, one can see  that the surface state contribution governs  the leading decay of
the critical Casimir interaction; it can be shown that  (\ref{eq:13}) as $x\to -\infty$ is $-(1/N^2)(\sinh 2K_1^*(T_c))^{-1} x^2e^{-x}$.
 At the bifurcation point  $x=-1$,  $(1/2)\gamma(\omega_1)$ goes to zero and is continuous
in its argument. The contribution coming  from the real mode $\omega_1$
is large and positive for $-1<x<0$  ( see blue dashed curve in the inset in Fig.~\ref{fig:1}) and has to be
 compensated in order to render the negative critical Casimir force in this interval. This contribution is given by (\ref{eq:13}) with 
$z$ replaced by $iz$, when now $z$ is  given by  a solution  of $\tan(z)=z/x$ for $-1<x<1$.
The Casimir force for $(++)$ BCs may be  obtained from the case of the free BCs by applying duality \cite{Roya}.
In this  case there is an asymptotic degeneracy associated with a surface mode {\it in the $T>T_c$ sector}.
Eqs.~(\ref{eq:13}) and (\ref{eq:14})
can be generalized to the case with the surface fields with $h_1$ and $h_2$ such that $h_1h_2>0$ \cite{we}.
This is an interesting matter because surface fields maybe adjusted to produce both attractive and repulsive Casimir forces.
Equally, considering the case $h_1h_2<0$ and scaling around the wetting transition but not the bulk critical point, we can produced novel behavior. These are rather detail calculations and results which seemed to the authors to make separate  publication advisable.

\vspace{2cm}

We now give a more intuitive account  of the asymptotic degeneracy.
For $N$ large enough, but finite, we will have states in the strip which are either largely positively or largely negatively magnetized. 
At the transfer matrix level, called these states $\mid\oplus\rangle$  and $\mid\ominus\rangle$.
We can choose the phases so that  the parity  operator   $P_N\mid\oplus\rangle = \mid\ominus\rangle$ 
(The  operator $P_N$ is given by  Eq.~(\ref{eq:6a}) with  $n=N$; note that
$P^2_N=1$. It is called
the parity operator because  it reverses all $x$-quantised spins in the column simultaneously; it is 
an invariant of the Hamiltonian.) 
 Then the fact that $P_N \mid\Phi\rangle =\mid\Phi \rangle$ implies
that $\mid\Phi\rangle =\left(\mid\oplus\rangle + \mid\ominus\rangle\right)/\sqrt2$ and further that  
$X^{\dagger}(iv(N))\mid\Phi\rangle=\left(\mid\oplus\rangle - \mid\ominus\rangle\right)/\sqrt2$, 
since $P_NX^{\dagger}(iv(N))\mid\Phi\rangle = - \mid\Phi\rangle$. It follows that 
\begin{equation}
 \label{eq:15}
\mid \oplus \rangle = \frac{(1+ X^{\dagger}(iv))\mid \Phi \rangle}{\sqrt2}\quad
 \mid \ominus \rangle = \frac{(1- X^{\dagger}(iv))\mid \Phi \rangle}{\sqrt 2} 
\end{equation}
The Euclidean ``evolution'' of these states is instructive: consider the matrix elements
$\langle\oplus\mid V^n\mid\oplus \rangle = (1+e^{-n\gamma(iv(N))})/2$  and
$\langle\ominus\mid V^n\mid\oplus\rangle = (1-e^{-n\gamma(iv(N))})/2$.
Now $\gamma(iv(N))\sim \hat\gamma(0) \exp(-N\hat\gamma(0))$ and if $n\hat\gamma(0)\gg \exp(N\hat\gamma(0))$, 
the transition matrix element tends to $1/2$, that is, the system dephases on this length scale. Thus 
 the dominant configurations correspond to successive postively and negatively magnetized regions separated by  domain walls running across
 the strip as in the Fisher Privman theory (see Fig.~1c in Ref.~\cite{privmann_fisher}). On the other hand, if $n=1$,
 then $\langle\ominus\mid V^n\mid\oplus\rangle \sim \gamma(iv)/2$ so that magnetic domain reversal is improbable but strictly not impossible if
$N<\infty$. Since $\langle\oplus\mid V^n\mid\oplus\rangle = \langle\ominus\mid V^n\mid\ominus\rangle$, we see that the wall interactions of the $\oplus$
and $\ominus$ phases are the same, an intuitively obvious result of symmetry. This is the physical reason for the smallness of $\gamma(iv(N))$
in (\ref{eq:10}). In (\ref{eq:6}), with $\omega_1=iv(N)$ we have $y_{2n-1} \sim N_0\exp(-nv(N))$ and $y_{2n} \sim iN_0\exp(-(N-n)v(N))$
with $v(N)\simeq \hat\gamma(0)$ ($N_0$ is a suitable normalization constant). The imaginary mode qualifies as a surface state
because  of the exponential attenuation away from the surface. This is because  $\hat\gamma(0)>0$, except at the critical point, which has a new attribute: it is also a de-pinning
point for the surface state, simultaneously at both edges as $\hat\gamma(0)\to 0$. In the scaling regime,
 the two exponentially decaying tails
overlap, a physical reason to expect a significant contribution to the Casimir force. The bubble/domain wall, or interface Hamiltonian, idea
has proved extremely useful in understanding planar uniaxial models \cite{abraham:83,AL,ASU}. It, too, needs amplification to 
work with free boundaries: for modes with real $\omega_j(N)$, path fluctuations are controlled by surface stiffness and by repulsion from the edges, as expected.
For imaginary mode we have a random succession of domain walls which cross the strip, pointing on average in the $(0,1)$ direction following Fisher-Privman.
Within a stripe with positive magnetization, the state in a column is $\mid\oplus\rangle$ and it is essentially proved that
the expectation value $\langle\oplus\mid\Gamma_{2n-1}\mid\oplus\rangle=y_{2n-1}$. Now $\Gamma_{2n-1}$ consists of a string of disorder
operators (the so-called the Jordan-Wigner tail see (\ref{eq:6a})), thus it creats a column of misfit bonds
 with which a path may be associated. This path connects the end of the tail at the site $n$ to any point on the boundary.
Thus $y_{2n-1}\sim \exp(-n\tau(0,1))$ where $\tau(0,1)=\hat\gamma(0)=v(N)$
is the surface tension in the $(0,1)$ direction.

In summary, in this Letter we have obtained the following new results:
(1.) Strong asymmetry of the Casimir force, as shown in Fig.~\ref{fig:1}, is related to the appearance 
in the spectrum of the transfer matrix of bound states, which are also surface states.
(2.) Surface states produce the asymptotic degeneracy of the transfer matrix advocated as a diagnostic for phase transition
 by Kac \cite{kac}. They also play a key role in the realization of Fisher-Privman theory in these systems. 
The critical point has an additional, new characterisation as an unbinding transition.
 We interpret the bound state at a microscopic level as a linear combination of block spin reversal operators.
(3.) Our results require a significant modification of the interface Hamiltonian concept for these systems. 

Occurence of  surface states depends on the boundary conditions of the strip.
In a strip with free boundaries  the asymptotic degeneracy
occurs because the coexistence of two pseudophases in a finite system remains at the same line 
in the thermodynamic space (vanishing bulk field $h=0$) as in the bulk system.
Surface states form because free boundaries effectively attract  domain walls due to the missing neighbours effect.
This is analogous to the case of surfaces with wetting boundary conditions below $T_c$, where
 the effective attraction due to the missing neighbours is amplified by the action of the surface fields. As a result, 
 the thin film can be formed near the surface which then undergoes unbinding at the wetting transition.
This argument carries over to three dimensional systems.
For $(+,+)$ boundary conditions an imaginary wavenumber and the associated  asymptotic degeneracy  occurs above $T_c$; 
there is no asymptotic degeneracy below $T_c$ 
because the surface breaks the symmetry 
and pseudo coexistence is shifted away from  the  $h=0$ line. 
For this boundary conditions, the imaginary mode gives rise to the critical adsorption which is characterized by the magnetization profile
which is enhanced near both surfaces and decays to zero in the middle of the strip.

\acknowledgments One of the author (DBA) acknowledges the support of Prof. S. Dietrich at the MPI Stuttgart 
and of the Center for Non-linear Studies at  Los Alamos
National Lab, where parts of this work were done.

\end{document}